\documentclass{iaus}
\usepackage{graphicx}

\title[JD 11.~~The origin of young local associations] 
{Using ages and kinematic traceback: \\ the origin of young local associations}

\author[D.Fern\'andez, F.Figueras \& J.Torra] 
{David Fern\'andez, Francesca Figueras
 \and Jordi Torra}

\affiliation{Departament d'Astronomia i Meteorologia, IEEC-Universitat de
Barcelona \\ Av. Diagonal 647, E-08028 Barcelona, Spain 
\\ email: {\tt david.fernandez@am.ub.es}}

\pubyear{2009}
\volume{xxx}  
\pagerange{1--8}
\jname{The Ages of Stars}
\editors{A.C. Editor, B.D. Editor \& C.E. Editor, eds.}
\begin{document}

\maketitle

\begin{abstract}
Over the last decade, several groups of young (mainly low-mass) stars have 
been discovered in the solar neighbourhood (closer than $\sim100$ pc), thanks to 
cross-correlation between X-ray, optical spectroscopy and kinematic data. 
These young local associations - including an important fraction whose members 
are Hipparcos stars - offer insights into the star formation process in 
low-density environments, shed light on the substellar domain, and could have 
played an important role in the recent history of the local interstellar 
medium. Ages estimates for these associations have been derived in the 
literature by several ways (HR diagram, spectra, Li and H$\alpha$ widths, 
expansion motion, etc.). In this work we have studied the kinematic evolution 
of young local associations and their relation to other young stellar groups 
and structures in the local interstellar medium, thus casting new light on 
recent star formation processes in the solar neighbourhood. We compiled the 
data published in the literature for young local associations, including the 
astrometric data from the new Hipparcos reduction. Using a realistic Galactic 
potential we integrated the orbits for these associations and the Sco-Cen 
complex back in time. Combining these data with the spatial structure of the 
Local Bubble and the spiral structure of the Galaxy, we propose a recent 
history of star formation in the solar neighbourhood. We suggest that both 
the Sco-Cen complex and young local associations originated as a result of 
the impact of the inner spiral arm shock wave against a giant molecular cloud. 
The core of the giant molecular cloud formed the Sco-Cen complex, and some 
small cloudlets in a halo around the giant molecular cloud formed young local 
associations several million years later. We also propose a supernova in young 
local associations a few million years ago as the most likely candidate to 
have reheated the Local Bubble to its present temperature.
\end{abstract}

\firstsection 
\section{Introduction}

In this work we propose a scenario for the history of the recent star
formation (during the last 20-30 Myr) in the nearest solar neighbourhood
($\sim$150 pc), from the study of the spatial and kinematic properties of
the members of the so-called young local associations, the Sco-Cen complex
and the Local Bubble, the most important structure observed in the local
interstellar medium (ISM).

\section{The local interstellar medium: the Local Bubble}

Locally, within the nearest 100 pc, the ISM is dominated by the Local Bubble
(LB).  The
displacement model (Snowden et al. \cite{Snowden98}) for
this structure assumes that the irregular local HI cavity is filled by an
X-ray-emitting plasma, with an emission temperature of $\sim10^6$ K and a
density, $n_e$, of $\sim 0.005$ cm$^{-3}$. Snowden et al.  
(\cite{Snowden98}) derived an extension for the LB of 40 to 130 pc, it
being larger at higher Galactic latitudes and smaller nearer the equator.
Lallement et al. (\cite{Lallement03})  
obtained the contours of the LB from NaI absorption measurements, tracing them 
with an estimated precision of
$\approx\pm$20 pc in most directions.

Several models have been presented to explain the origin of the LB. The 
consensus is reached with a scenario in which about 10-20 supernovae (SNe) 
formed the local cavity and, after that, a few SNe reheated the LB a few Myr 
ago, explaining the currently observed temperature of the diffuse soft X-ray 
background (Breitschwerdt
\& Cox \cite{Breitschwerdt04}). Some authors have remarked that there is
independent evidence for the occurrence of a close SN ($\sim$30 pc)
$\sim$5 Myr ago (Knie et al. \cite{Knie99}), which could be the best
candidate for reheating the LB. Ma\'{\i}z-Apell\'aniz (\cite{MaizApellaniz01}) 
proposed that the 2 or 3 SNe that reheated the local cavity could have 
exploded in LCC (one of the OB associations of the Sco-Cen complex), but his 
results faced some geometrical problems due to the 
peripheral situation of LCC with respect to the LB.

\section{Local associations of young stars}

A decade ago, very few PMS stars had been identified less than 100 pc from
the Sun. Nearly all the youngest stars ($\le$30 Myr) studied then were
located more than 140 pc away in the molecular clouds of Taurus,
Chamaeleon, Lupus, Sco-Cen and R CrA. The cross-correlation of the Hipparcos 
and ROSAT catalogues suddenly changed this; a few stars were identified as 
very young but closer than 100 pc, where there are no molecular clouds with 
stellar forming regions (Neuh\"auser \& Brandner \cite{Neuhauser98}). 
Two explanations for the existence of these
young stars far away from SFR were proposed. Sterzik \& Durisen
(\cite{Sterzik95}) suggested that the stars were formed in molecular
clouds and later ejected as high-velocity stars during the decay of young
multiple star systems. Feigelson (\cite{Feigelson96}) suggested that the
stars were formed inside small molecular clouds (or {\it cloudlets}), which
later dispersed among the ISM and therefore can no longer be detected.

\begin{figure}[b]
\begin{center}
 \includegraphics[width=10cm]{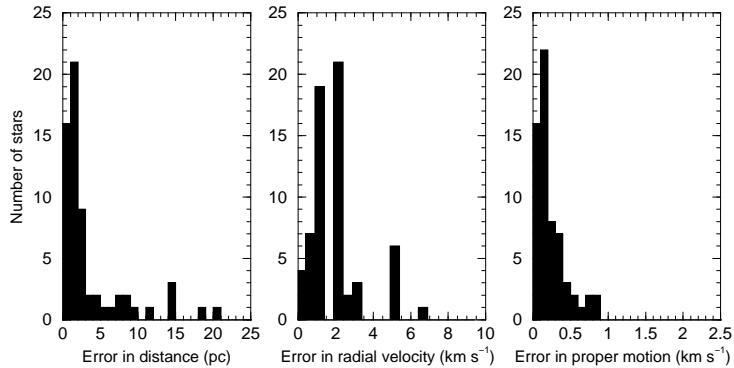}
 \caption{Observational errors in distance (left), radial velocity (center) and
          tangential velocity (right) for the stars of our compilation of YLA.}
   \label{fig.Error.R+mu}
\end{center}
\end{figure}

These young nearby stars were grouped into clusters, associations and
moving groups, each with a few dozen members. Different approaches were
used in each YLA discovery, but most of them made use of Hipparcos proper
motions, X-ray emission, infrared emission and ground-based spectroscopy 
and photometry. We have compiled all the published YLA data. Astrometric data 
come from
the new Hipparcos reduction (van Leeuwen \cite{vanLeeuwen07}). 
Table \ref{tab.Assoc} 
shows the mean spatial and kinematic
properties, ages and number of members for each YLA. The adopted age shown in 
Table \ref{tab.Assoc} is that assigned for back-tracing the association
orbits in the next section. In Fig. \ref{fig.Error.R+mu} the observational
errors for the stars of our compilation are shown.

\begin{table}
\begin{center}
   \caption{Mean spatial coordinates and heliocentric velocity components
            of the young local associations and the Sco-Cen complex (in
            the latter case, data from de Zeeuw \cite{deZeeuw99} (Z99), 
	    Madsen et al. \cite{Madsen02} (M02) and Sartori et al.
	    \cite{Sartori03} (S03)).
            In brackets, the standard deviation of the sampling
            distribution. $N$ is the number of known members in each
            association ($N_\mathrm{k}$ with complete kinematic data).}
   \label{tab.Assoc}
{\scriptsize
\begin{tabular}{lrrrrrrrrrrr}
\hline
\hline
Association     & $\overline{\xi^\prime}$ &
                  $\overline{\eta^\prime}$ &
                  $\overline{\zeta^\prime}$
                & $\overline{r}$
                & $\overline{U}$ & $\overline{V}$ & $\overline{W}$
                & Age & $N$ & $N_\mathrm{k}$ \\
                & (pc) & (pc) & (pc) & (pc)
                & (km s$^{-1}$) & (km s$^{-1}$) & (km s$^{-1}$) & (Myr)
                & & \\
\hline
TW Hya          & $-18_{(12)}$
                & $-51_{(16)}$
                & $22_{(\;\;7)}$
                &  $60_{(18)}$
                &  $-8.4_{(3.9)}$ & $-18.3_{(3.4)}$ &  $-4.9_{(1.8)}$
                &  8 & 39 & 5 \\
Tuc-Hor/GAYA    & $-12_{(26)}$
                & $-26_{(12)}$
                & $-36_{(13)}$
                &  $53_{(17)}$
                & $-11.4_{(7.3)}$ & $-22.4_{(5.5)}$ &  $-2.7_{(3.7)}$
                &  20 & 52 & 28 \\
$\beta$ Pic-Cap & $-10_{(26)}$
                & $-7_{(13)}$
                & $-14_{(10)}$
                &  $37_{(13)}$
                & $-10.1_{(3.5)}$ & $-15.2_{(3.7)}$ &  $-10.2_{(2.9)}$
                &  12 & 33 & 21 \\
$\epsilon$ Cha  & $-52_{(\;\;3)}$
                & $-89_{(\;\;5)}$
                & $-27_{(\;\;2)}$
                &  $107_{(\;\;6)}$
                &  $-10.2_{(0.8)}$ & $-18.9_{(0.6)}$ &  $-10.2_{(2.0)}$
                &  10 & 16 & 4 \\
$\eta$ Cha      & $-31\;\;\;\;\;\;$
                & $-75\;\;\;\;\;\;$
                & $-32\;\;\;\;\;\;$
                &  $88\;\;\;\;\;\;$
                & $-10.1\;\;\;\;\;\;\;$ & $-18.4\;\;\;\;\;\;\;$ & $-10.0\;\;\;\;\;\;\;$
                &  10 & 18 & 1 \\
HD 141569       & $-77_{(\;\;3)}$
                & $10_{(\;\;8)}$
                & $64_{(\;\;8)}$
                & $101_{(\;\;8)}$
                &  $-5.4_{(1.5)}$ & $-15.6_{(2.6)}$ &  $-4.4_{(0.8)}$
                &  5 & 5 & 2 \\
Ext. R CrA      & $-81_{(33)}$
                & $-6_{(\;\;6)}$
                & $-28_{(\;\;9)}$
                & $87_{(33)}$
                &  $-2.2_{(6.1)}$ & $-15.8_{(1.4)}$ & $-10.9_{(0.8)}$
                &  13 & 59 & 2 \\
\hline
US \hspace{0.53cm}Z99 & $-141_{(34)}$ & $-22_{(11)}$ & $50_{(16)}$
                     & $145_{(\;\;2)}$
                     & & &
                     & 5-6 & 120 &  \\
\hspace{1.0cm}M02$^1$ & $-138_{(27)}$ & $-22_{(10)}$ & $49_{(12)}$
                     & $149_{(28)}$
                     & $-0.9\;\;\;\;\;\;\,$ & $-16.9\;\;\;\;\;\;\,$ &
		     $-5.2\;\;\;\;\;\;\,$
                     & & 120 &  &  \\
\hspace{1.0cm}S03    & & & &
                     & $-6.7_{(5.9)}$ & $-16.0_{(3.5)}$ & $-8.0_{(2.7)}$
                     & 8-10 & 155 &  \\
\hline
UCL \hspace{0.3cm}Z99 & $-122_{(30)}$ & $-69_{(26)}$ & $32_{(16)}$
                     & $140_{(\;\;2)}$
                     & & &
                     &  14-15 & 221 &  \\
\hspace{1.0cm}M02$^1$ & $-121_{(26)}$ & $-68_{(21)}$ & $32_{(15)}$
                     & $145_{(24)}$
                     & $-7.9\;\;\;\;\;\;\,$ & $-19.0\;\;\;\;\;\;\,$ & $-5.7\;\;\;\;\;\;\,$
                     & & 218 &  &  \\
\hspace{1.0cm}S03    & & & &
                     & $-6.8_{(4.6)}$ & $-19.3_{(4.7)}$ & $-5.7_{(2.5)}$
                     & 16-20 & 262 &  &  \\
\hline
LCC \hspace{0.3cm}Z99 & $-62_{(18)}$ & $-102_{(24)}$ & $14_{(16)}$
                     & $118_{(\;\;2)}$
                     & & &
                     &  11-12 & 180 &  \\
\hspace{1.0cm}M02$^1$ & $-61_{(14)}$ & $-100_{(15)}$ & $14_{(15)}$
                     & $120_{(18)}$
                     & $-11.8\;\;\;\;\;\;\,$ & $-15.0\;\;\;\;\;\;\,$ & $-6.7\;\;\;\;\;\;\,$
                     & & 179 &  \\
\hspace{1.0cm}S03    & & & &
                     & $-8.2_{(5.1)}$ & $-18.6_{(7.3)}$ & $-6.4_{(2.6)}$
                     & 16-20 & 192 &  \\
\hline
\multicolumn{11}{l}{\tiny $^1$ M02 derived an internal velocity dispersion 
of 1.33 km s$^{-1}$ for US, 1.23 km s$^{-1}$ for UCL and 1.13 km s$^{-1}$
for LCC.} \\
\end{tabular}
}
\end{center}
\end{table}

\section{Ages and integration of orbits}

The integration back in time of the Sco-Cen and YLA orbits allows us to
study their origin and possible influence on the local ISM over the
last million years. To compute the stellar orbits back in time we used the
code developed by Asiain et al. (\cite{Asiain99}) based on the
integration of the equations of motion using a realistic model of the
Galactic gravitational potential. We decomposed this potential
into three components: the general axisymmetric potential
$\Phi_\mathrm{AS}$, the potential
due to the spiral structure of the Galaxy $\Phi_\mathrm{Sp}$, and that due
to the central bar $\Phi_\mathrm{B}$. Details on the method can be found in 
Fern\'andez et al.
(\cite{Fernandez08}).

\subsection{Orbits of individual stars: are we able to derive dynamic ages?}

One of the most interesting results that could be derived from the study of 
the past trajectories of the individual members of the young local 
associations is the direct determination of their age. This could be achieved 
if one observes a clear spatial concentration of the members of one specific 
association in the past. 

\begin{figure}[t]
\begin{center}
 \includegraphics[width=6cm]{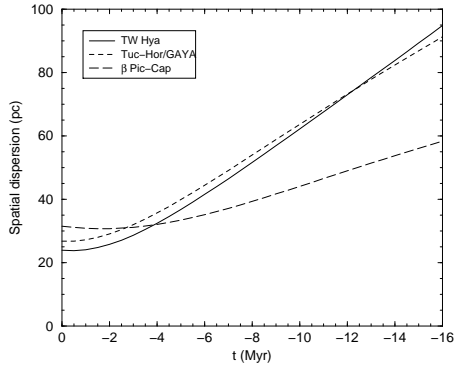}
 \caption{Spatial dispersion of those stars belonging to TW Hya, Tuc-Hor/GAYA 
          and $\beta$ Pic-Cap during the last Myr.}
   \label{fig.SpatialDispersion}
\end{center}
\end{figure}

However, we have not found this concentration in the past. We have computed 
the trajectories back in time for the stars of the three associations with the 
largest number of members with complete, high-quality data. In Fig.
\ref{fig.SpatialDispersion} we show the spatial dispersion of these stars as a 
function of time in the past, the results are those shown in the figure. As 
can be seen, during the last 4 million years, the spatial dispersion remains 
similar to that observed at present, but it grows rapidly to the past. So, it 
seems that the combination of the observational errors and the intrinsic 
velocity dispersion prevent us from deriving the ages for the associations.
Or maybe there are some problems with the membership assignment of the stars to
each association. Then, in our analysis we will only work with the past 
trajectories of the associations as a whole.

\subsection{The origin of young local associations}

\begin{figure}
\begin{center}
 \includegraphics[width=8cm]{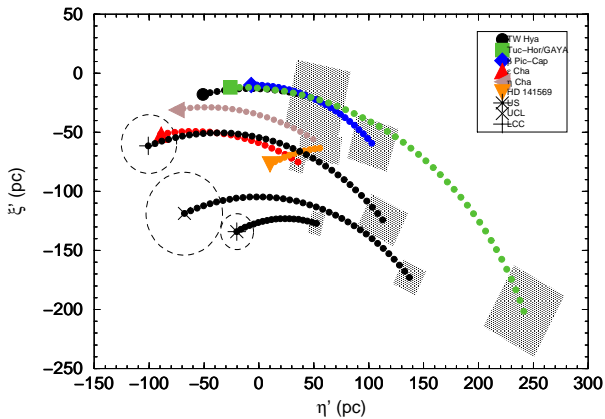}
 \caption{Positions and orbits in the Galactic plane ($\xi^\prime,\eta^\prime$) 
         of YLA and the Sco-Cen complex going back in time to their individual 
	 ages.
	 The grey areas show the expected positional errors at birth due
	 to kinematic observational errors. The centre of the ($\xi^\prime,
	 \eta^\prime$) coordinate system is comoving with the LSR.}
   \label{fig.Associacions+ScoCen.ErrorsCin}
\end{center}
\end{figure}

We have integrated the orbits of the associations using the mean
position and velocity for each association (Table \ref{tab.Assoc}). The
results are presented in Fig. \ref{fig.Associacions+ScoCen.ErrorsCin}. 
We show the estimated error in the position of the associations at birth (grey 
area). An error in age 
shall be read as a displacement of the error areas along the plotted orbits in 
the figure. The most obvious trend observed is the spatial concentration of 
all the associations (Sco-Cen complex and YLA) in the first Galactic quadrant 
in the past. There is a very 
conspicuous spatial grouping at the time of birth of TW Hya, $\epsilon$ Cha, 
$\eta$ Cha and HD 141569 (in a sphere 25 pc in radius). At its birth, $\beta$ 
Pic-Cap was located about 50 pc from the other YLA. The region where these 
associations were formed had a size of 
50 x 70 x 40 pc ($\xi^\prime$ x $\eta^\prime$ x $\zeta^\prime$). At present 
the volume has increased by a factor of $\sim3.5$. The errors associated with 
the mean 
velocity components of the associations do not have a crucial influence 
on the previous results: the error areas have typical side lengths of about 
10-30 pc.

Mamajek et al. (\cite{Mamajek00}) found that extrapolating past motions
(assuming linear ballistic trajectories) shows that TW Hya, $\epsilon$
Cha, $\eta$ Cha and the three subgroups of the Sco-Cen complex were
closest together about 10-15 Myr ago. They suggest that these three
associations were formed in the progenitor Sco-Cen GMC or in short-lived
molecular clouds formed by Sco-Cen superbubbles. Jilinski et al. 
(\cite{Jilinski05}) locate
the birthplaces of $\epsilon$ Cha and $\eta$ Cha at the edge of LCC.  
Ortega et al. (\cite{Ortega04}) suggest that the
$\beta$ Pic moving group was formed near Sco-Cen, probably due to a
SN in this complex. Our compendium of all the known members of the
whole set of YLA and the integration of their orbits back in time allow us
to present here a more detailed analysis of the origin of YLA.

\begin{figure}
\begin{center}
 \includegraphics[width=6cm]{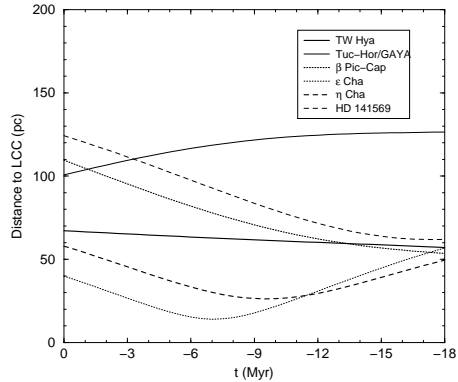}
 \caption{Distances between the centres of YLA
         and the centre of LCC as a function of 
         time, using the Sco-Cen kinematics in S03.}
  \label{fig.DistLCC}
\end{center}
\end{figure}

In Fig. \ref{fig.DistLCC} the temporal
evolution of the distances between each one of the YLA and the centre of LCC 
(the nearest OB association) is
shown. The instant when the distance minima between LCC and YLA occurred is of
great interest. In the cases of the $\eta$ Cha cluster and the $\epsilon$
Cha association, minima with distances of 14 and 26 pc, respectively, are
obtained for $t \sim -$(7-9) Myr. No clear
distance minima to LCC are found in the recent past ($t > -20$ Myr) for
the other YLA. TW Hya was 62 pc from LCC 8 Myr ago (the estimated age for
this association), whereas the HD 141569 system was about 102 pc from LCC 5
Myr ago, continuously decreasing to 62 pc 18 Myr ago. The Tuc-Hor/GAYA
association has maintained a distance to LCC of more than 100 pc over the
last 20 Myr, but this could be the only YLA considered here not to
originate from a SN in LCC, since its estimated age is equal to
(or even larger than) that derived for LCC.

We conclude that, at the moment of their birth, YLA were
at distances of between 15 and 100 pc from the {\it centre} of LCC, and
even further from the other two Sco-Cen associations. Although observational 
errors in parallax and velocity
components, as well as errors in age estimations, could affect these
results, the fact that we work with mean values for distances, velocity
components and ages minimises this possibility. Although the present radius of 
LCC is about 25 pc, it has been
continuously expanding since birth, 16-20 Myr ago. Even considering a
moderate expansion rate, one should expect an initial radius $\le$20
pc. The distances obtained from Fig. \ref{fig.DistLCC}, together with the 
expected reliability of the
orbits, lead us to believe that the local 
associations were not born inside the cloud that formed the Sco-Cen complex, 
but in small molecular clouds outside it.

One possible scenario for the formation of YLA in these small molecular clouds 
is the explosion of one or several close SNe, which could have produced 
compression that triggered star formation. These hypothetical SNe should 
belong to the Sco-Cen 
complex. This complex is made up 
of several thousand stars, more than 300 of which are early-type stars, and 
around 35 are candidates for Type II SNe. Ma\'{\i}z-Apell\'aniz 
(\cite{MaizApellaniz01}) 
estimated the number of past SNe inside the three associations, 
obtaining 1 SN 
for US, 13 for UCL and 6 for LCC. The first SN that exploded in each 
association took place when it was 3-5 Myr old, and the others have been 
exploding and will continue to explode at a nearly constant rate, for the first 
$\sim$30 Myr of the complex's life. Even a conservative estimate gives at 
least 6 SNe in UCL during the last 10-12 Myr, another 6 in LCC during 
the last 7-9 Myr and at least 1 in US.

The wave front of a SN typically moves at a velocity of a few tens
of pc per million years; therefore, a SN explosion in LCC or UCL 
9-11 Myr ago may have triggered star formation between 1 and 3 Myr later in
small molecular clouds at distances of 15 to 75 pc. These would be the
parent clouds of $\eta$ Cha and $\epsilon$ Cha, located at $\sim$20 pc from 
LCC at their birth. Taking into
account that the first SNe in LCC exploded when the
associations were 3-5 Myr old (Ma\'{\i}z-Apell\'aniz
\cite{MaizApellaniz01}), this scenario is only possible for a present age
of LCC of at least 12 Myr. This is not a problem, since the
estimated ages for LCC published in the literature range from 11-12 Myr to
16-20 Myr. If only one SN could explain this star formation outbreak
$\sim$8.5-9 Myr ago, this would be the age of $\eta$ Cha and $\epsilon$
Cha, which would have been formed simultaneously. It should be remembered
that the estimated ages for $\eta$ Cha and $\epsilon$ Cha are 5-15 and $\le$10 
Myr, respectively.

Such a SN could also have triggered the formation of TW Hya, whose
estimated age is $\sim$8 Myr. As mentioned above, at that time TW Hya was
about 45 pc from the centre of LCC, in perfect agreement with the typical
distance at which a SN wave front can trigger star formation in a
small molecular cloud. However, it was not necessarily a single SN
in LCC or UCL that was the origin of these four YLA. The SN rate in
these two associations is $\sim$0.5 Myr$^{-1}$ and, therefore, it is
possible that a few SNe in the period $-7 \le t \le -10$ Myr
triggered the star formation that resulted in YLA. In any case, from our
results we can conclude that these associations definitely did not form
inside the associations of Sco-Cen, to be later ejected. They were formed in regions of
space far from Sco-Cen, probably in small molecular clouds that
were later totally dispersed by the newly born stars and/or by the shock
fronts of later SNe in Sco-Cen.

Our results support a star formation scenario for very young stars far
away from SFR or molecular clouds, such as that proposed by Feigelson
(\cite{Feigelson96}) and not that of Sterzik \& Durisen
(\cite{Sterzik95}). The latter authors perform numerical simulations to
explain the existence of haloes of isolated T Tau stars around SFR. In
their simulations a significant number of stars were ejected from these
regions at birth with large velocities, allowing trajectories of some tens
of pc in a few million years. Meanwhile, Feigelson (\cite{Feigelson96})
proposed another scenario for the formation not only of the haloes of T
Tau stars, but also of other completely isolated very young stars that
have been discovered. In this model, the isolated T Tau stars form in
small, fast-moving, short-lived molecular clouds. The gas remaining after
the star formation process is rapidly dispersed by the stellar winds of
the new stars. At present the stars are located in regions of space where
there is no gas and so, apparently, they have formed far away from any
SFR. The case of YLA supports this scenario, since our kinematic study
shows that these associations formed far away from the Sco-Cen complex. For 
the HD 141569 system, a SN in UCL, as opposed to one in LCC, is a more 
promising candidate to explain its origin. This is because the distance to LCC 
for the range of ages accepted for this group (2-8 Myr) is between 88 and 116 
pc, whereas for UCL it is 72-88 pc.

\subsection{Young local associations and Local Bubble}

If we superimposed on the present LB structure the trajectories back
in time for YLA (see Fig. 13 in Fern\'andez et al. \cite{Fernandez08})
we can see that the past orbits of the YLA are closer to the central region 
of the LB than Sco-Cen associations. To 
be exact, the trajectories of the centres of the associations TW Hya, 
Tuc-Hor/GAYA and $\beta$ Pic-Cap have crossed very near to the geometric centre 
of the LB in the last $\sim$5 Myr. The uncertainty boxes in position on the 
Galactic plane do not exceed a 
few tens of pc. On the other hand, errors in age estimates for the YLA result 
in positional errors along the trajectories back in time. Even considering the 
large uncertainties in age obtained for some of these YLA, ages are not 
expected to exceed 20 Myr for any of them 
(except for Tuc-Hor/GAYA). We can therefore conclude that the YLA have been 
moving inside the LB for (at least) most of their lifetime, and can question 
whether the presence of these young stars inside the LB bears any relation to 
its origin and/or evolution.

All the YLA except TW Hya and $\beta$
Pic-Cap contain B-type stars (13 stars from a total of 223 members; see Table 6
in Fern\'andez et al. \cite{Fernandez08}). It is not possible to derive the 
number of stars earlier than B2.5
that were born in the local associations, since we do not know their
total mass precisely. However, the fact that at present we observe one
SN candidate (a B2IV star), and more than a dozen stars of spectral 
types between B5 and B9, allows us to affirm that it is possible that one or
more of these associations has sheltered a SN in the recent past
(the last 10 Myr). As there is direct evidence for an explosion of a
SN at a distance of $\sim$30 pc, $\sim$5 Myr ago, several
pieces of the same puzzle seem to support the theory of a recent SN
in the nearest solar neighbourhood originating from a parent star
belonging to a YLA, probably Tuc-Hor/GAYA or the extended R CrA
association (which currently show the highest content of B-type stars).  
This near and recent SN would have been responsible for the
reheating of the gas inside the LB needed to achieve the currently observed
temperature of the diffuse soft X-ray background.

As we mention above, there is no agreement in the literature on the number
of SNe needed to form the local cavity. If only one was enough, the 
SN we propose would be the most promising candidate, since it
would be placed very near the geometric centre of the LB, explaining in a
natural way its present spatial structure. If more SNe are needed
(as recent works suggest), we could consider other stars in the vicinity
of LCC and UCL, as proposed by Fuchs et al. (\cite{Fuchs06}).

\section{A scenario for the local and recent star formation}

\begin{figure}[t]
\begin{center}
 \includegraphics[width=8cm]{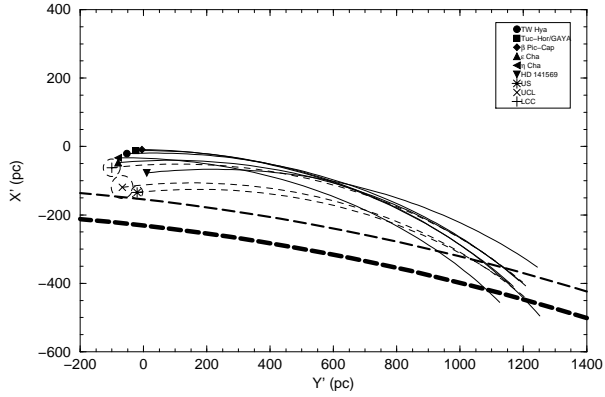}
 \caption{Orbits in the Galactic plane  
         ($X^{\prime}$-$Y^{\prime}$) integrated back in time
         to $t = -$30 Myr for YLA
         and the three associations of the Sco-Cen complex. 
         The thick-dashed line 
         shows the position of the minimum of the spiral potential 
         (Fern\'andez et al. \cite{Fernandez01}). The thin-dashed line is the 
         position of the phase of the spiral structure $\psi = 10^o$.}
   \label{fig2}
\end{center}
\end{figure}

If the impact of the spiral arm shock wave was the initial cause of star
formation in the Sco-Cen region (see details in Fern\'andez et al.
\cite{Fernandez08}), then
the history of the nearest solar neighbourhood during the last few tens of
Myr would have been as follows. 30 Myr ago the GMC that
became the parent of Sco-Cen was in the Galactic plane with coordinates
$(X^{\prime},Y^{\prime}) \sim (-400,1200)$ pc (see Fig.  
\ref{fig2}). The arrival of the
potential minimum of the inner spiral arm triggered star formation in the
region. At the same time it disturbed the cloud's motion, whose velocity
vector became directed in the opposite direction to Galactic rotation and
away from Galactic centre. 
The compression due to the spiral arm
did not necessarily trigger star formation in the whole cloud, but perhaps
only in the regions with the largest densities. This would be favoured by
the smaller relative velocity between the shock wave and the RSR. 
The regions where star formation began must be
those which generated UCL, LCC and, probably, the Tuc-Hor/GAYA
association, which were all born at nearly the same time: about 16-20 Myr
ago. The stellar winds from the first massive stars began to compress the
gas of the neighbouring regions, maybe causing them to fragment into small
molecular clouds that moved away from the central region of the parent
cloud. About 9 Myr ago, a SN in LCC or UCL triggered star formation
in these small molecular clouds, giving birth to the majority of YLA, as
we saw in the previous section. The stellar winds of the newly born stars
rapidly expelled the remaining gas from these small clouds (the {\it
cloudlets} proposed by Feigelson \cite{Feigelson96}), completely erasing
every trace of them and leading to our observation that there is no gas in
these regions at present. YLA may have had a crucial influence on the
history of the LB. We suggest that one or two SNe in these associations
were responsible for reheating the LB a few million years ago. This hypothesis
seems to be reinforced by the evidence of a very near SN about 5
Myr ago (Knie et al. \cite{Knie99}). At about the same time, as proposed
by Preibisch \& Zinnecker (\cite{Preibisch99}), the shock front of a
SN in UCL would have triggered star formation in US about 6 Myr
ago. Only 1.5 Myr ago, the most massive star in US would have gone
SN and its shock front would now be reaching the molecular cloud of
$\rho$ Oph, triggering the beginning of the star formation process there.

\begin{discussion}

\discuss{Melbourne}{What effect do these SNe have on the Earth? Are there 
tracers on Earth that suggest recent SNe explosions?}

\discuss{Fern\'andez}{In a paper published in 1999 by Knie and collaborators, 
they claimed to have found in a deep ocean crust evidences of the explosion of 
a SN in the recent past. They proposed that this SN exploded 5 
Myr ago at a distance of about 30 pc.}

\discuss{Mamajek}{Other researchers have calculated dynamical ages based on 
the time of minimum pass between some groups in the past. We are all here 
because we want improved ages for stars. How useful do you think these 
dynamical ages are?}

\discuss{Fern\'andez}{We have found that deriving dynamic ages is not a 
reliable aging methodology, al least considering the present membership 
assignment for each one of the associations. Other authors have derived 
dynamic ages, but using what they called "core stars"; that is, selecting 
those stars with the minimum dispersion in their velocity components. I 
strongly believe that the present membership assignment must be revised to 
be able to derive reliable dynamic ages for each one of these young local 
associations.} 

\discuss{Naylor}{You have a diagram which shows the motion of the local 
associations after they passed through the spiral arm. Where does the Sun 
move on this diagram? How long have we been near the local associations?}

\discuss{Fern\'andez}{We have not computed the orbit of the Sun in this 
figure, so I do not know exactly how long the Sun has been near the young 
local associations.}

\discuss{Roberto}{Are the counts of low-mass stars in local associations 
compatible with the number of SNe?}

\discuss{Fern\'andez}{The problem with the counts of low-mass stars is that 
they are not complete and it is difficult to say if these counts are 
compatible with the number of SNe we propose (one or two). However, the counts 
of early-type stars, which are complete, seems to support the hypothesis that 
a SN exploded in the recent past in the young local associations.} 

\end{discussion}

\end{document}